\documentclass[english,12pt,journal, draftclsnofoot,onecolumn]{IEEEtran}
\usepackage[T1]{fontenc}
\usepackage[latin9]{inputenc}
\usepackage{color}
\usepackage{amsmath}
\usepackage{amsthm}
\usepackage{amssymb}
\usepackage{graphicx}
\usepackage{esint}

\makeatletter

\newcommand{\lyxdot}{.}

\theoremstyle{plain}
\newtheorem{thm}{\protect\theoremname}
\theoremstyle{remark}
\newtheorem{rem}[thm]{\protect\remarkname}
\theoremstyle{plain}
\newtheorem{prop}[thm]{\protect\propositionname}

\usepackage{psfrag,color,cite}
\usepackage{tikz,pgfplots}
\usepackage{paralist}
\usepackage[caption=false]{subfig}
\usepackage{flushend}
\usepackage{url}

\@ifundefined{showcaptionsetup}{}{%
 \PassOptionsToPackage{caption=false}{subfig}}
\usepackage{subfig}
\makeatother

\usepackage{babel}
\providecommand{\propositionname}{Proposition}
\providecommand{\remarkname}{Remark}
\providecommand{\theoremname}{Theorem}

\begin{document}

\title{Packet Reception Probabilities in Vehicular Communications Close
to Intersections}

\author{Erik Steinmetz~\IEEEmembership{Student Member,~IEEE}, Matthias
Wildemeersch~\IEEEmembership{Member,~IEEE}, Tony Q.S. Quek,~\IEEEmembership{Senior Member,~IEEE},
and Henk Wymeersch,~\IEEEmembership{Member,~IEEE}\thanks{E. Steinmetz and H. Wymeersch are with the Department of Signals and
Systems, Chalmers University of Technology, Gothenburg, Sweden, e-mails:
\{estein,henkw\}@chalmers.se. E. Steinmetz is also with SP Technical
Research Institute of Sweden,  Borås, Sweden. M. Wildemeersch is with
the International Institute for Applied Systems Analysis (IIASA),
Laxenburg, Austria, e-mail: wildemee@iiasa.ac.at. T.Q.S. Quek is with
Singapore University of Technology and Design, Singapore, e-mail:
tonyquek@sutd.edu.sg. This research was supported, in part, by the
European Research Council under Grant No.~258418 (COOPNET), the EU
project HIGHTS (High precision positioning for cooperative ITS applications)
MG-3.5a-2014-636537, and VINNOVA under the program ``Nationell Metrologi
vid SP Sveriges Tekniska Forskningsinstitut''. Part of this work
was presented in \cite{Steinmetz2015}. }}
\maketitle
\begin{abstract}
Vehicular networks allow vehicles to share information and are expected
to be an integral part in future intelligent transportation system
(ITS). In order to guide and validate the design process, analytical
expressions of key performance metrics such as packet reception probabilities
and throughput are necessary, in particular for accident-prone scenarios
such as intersections. In this paper, we analyze the impact of interference
in an intersection scenario with two perpendicular roads using tools
from stochastic geometry. We present a general procedure to analytically
determine the packet reception probability and throughput of a selected
link, taking into account the geographical clustering of vehicles
close to the intersection. We consider both Aloha and CSMA MAC protocols,
and show how the procedure can be used to model different propagation
environments of practical relevance. We show how different path loss
functions and fading distributions can be incorporated in the analysis
to model propagation conditions typical to both rural and urban intersections.
Our results indicate that the procedure is general and flexible to
deal with a variety of scenarios. Thus, it can serve as a useful design
tool for communication system engineers, complementing simulations
and experiments, to obtain quick insights into the network performance.
\end{abstract}

\section{Introduction\label{sec:Introduction}}

\IEEEPARstart{V}{ehicular} networks have gained considerable attention
in the past years and are regarded as one of the key components in
future intelligent transportation systems~(ITS) \cite{Karagiannis2011,Papadimitratos2009,Hartenstein2008,Dar2010,Anjum2007,Alsabaan2013}.
By the use of wireless communication they allow vehicles to continuously
share information with each other and their surrounding (e.g., roadside
infrastructure), in order to perceive potentially dangerous situations
in an extended space and time horizon \cite{Papadimitratos2009}.
This enables a new set of applications that are expected to enhance
both traffic safety and efficiency. These applications include lane
change assistance, cooperative collision avoidance, emergency vehicle
warning, traffic condition warning, tolling, hazardous location warning,
speed management. 

The IEEE~802.11p standard has been defined to meet the communication
demand of these applications, and 5G cellular networks standards are
being developed to support device-to-device~(D2D) communication\cite{Cheng2007,Monserrat2014,Mumtaz2014,Yilmaz2014,Khelil2014,Araniti2013}.
However, different ITS applications clearly have different requirements
on the communication links, with the most stringent demands imposed
by safety-related applications, with extremely low latencies (below
50 ms in pre-crash situations), high delivery ratios (for full situational
awareness), and relatively long communication ranges (to increase
the time to react in critical situations)\cite{Johri2016,Santa2010,Mecklenbrauker2011}.
These requirements, in combination with a possible high density of
vehicles, makes the design of vehicular communication systems challenging.
This is further exacerbated by high mobility and passing vehicles,
which leads to rapidly changing signal propagation conditions (including
both severe multipath and shadowing) and constant topology changes.

In order to guide and validate the communication system design, extensive
simulations and measurements are often used \cite{Mecklenbrauker2011,Sjoberg2013},
which are both time consuming and scenario-specific. In order to obtain
insight in scalability and performance, analytical expressions of
key performance metrics are necessary, in particular for high velocity
scenarios (in particular highways) and accident-prone scenarios (e.g.,
intersections). Stochastic geometry is a tool to obtain such expressions,
and has been widely used in the design and analysis of wireless networks
\cite{Haenggi2008}. In 2-D planar networks, the analysis is well
developed and a multitude of approaches to consider both geographical
and medium access control~(MAC) induced clustering \cite{Ganti2009,Deng2014}
as well as different types of fading \cite{Hunter2008,Heath2012,Blaszczyszyn2013}
exist. However, in vehicular networks, where the location of the nodes
are restricted by the roads, previous work that includes the spatial
statistics of vehicles typically considers one-dimensional roads \cite{Jeong2013,Tong,Blaszczyszyn2009,Baszczyszyn2012,Blaszczyszyn2012}.
For these vehicular scenarios, geographical clustering has been addressed
in \cite{Jeong2013}, while effects due to the 802.11p carrier sense
multiple access~(CSMA) MAC protocol was studied in \cite{Tong,Blaszczyszyn2009,Nguyen2011}.
Hence, these works enable communication system design for one-dimensional
highway scenarios, but do not capture well the salient effects of
intersections. Intersections were considered explicitly in \cite{steinmetz2014communication,Steinmetz2015},
which found that it is important to properly model the interference
from different roads and account for the distance of receivers to
the intersection, i.e., to take into account the clustering of cars
around the intersection and the non-stationarity of the spatial distribution.

In this paper, we present a general procedure for the evaluation of
packet reception probability and throughput in intersection scenarios,
and provide a model repository that can be used to adapt to a variety
of different environments of importance in the vehicular context.
This includes both rural and urban scenarios, different propagation
conditions, and different MAC protocols. Latency and mobility are
not treated in this paper. 

The remainder of the paper is organized as follows. Section~\ref{sec:System-Model}
introduces the system model. In Section~\ref{sec:Typical-Scenarios},
we discuss typical characteristics of the vehicular channel and show
how the model can be tailored to different environments. In Section~\ref{sec:Stochastic-Geometry-Analysis},
we present a general procedure to calculate the packet reception probability
near an intersection, as well as the throughput. Section~\ref{sec:Case-Studies}
shows how the proposed procedure can be used to calculate these performance
metrics for a number of cases of practical relevance, and how different
assumptions on loss function, fading, and MAC protocols affect the
analytical tractability of the packet reception probability. Finally,
Section~\ref{sec:Conclusions} summarizes and concludes the paper.

\section{System Model\label{sec:System-Model}}

\begin{figure}[h]
\center

\subfloat[intersection scenario]{\includegraphics[width=0.4\columnwidth]{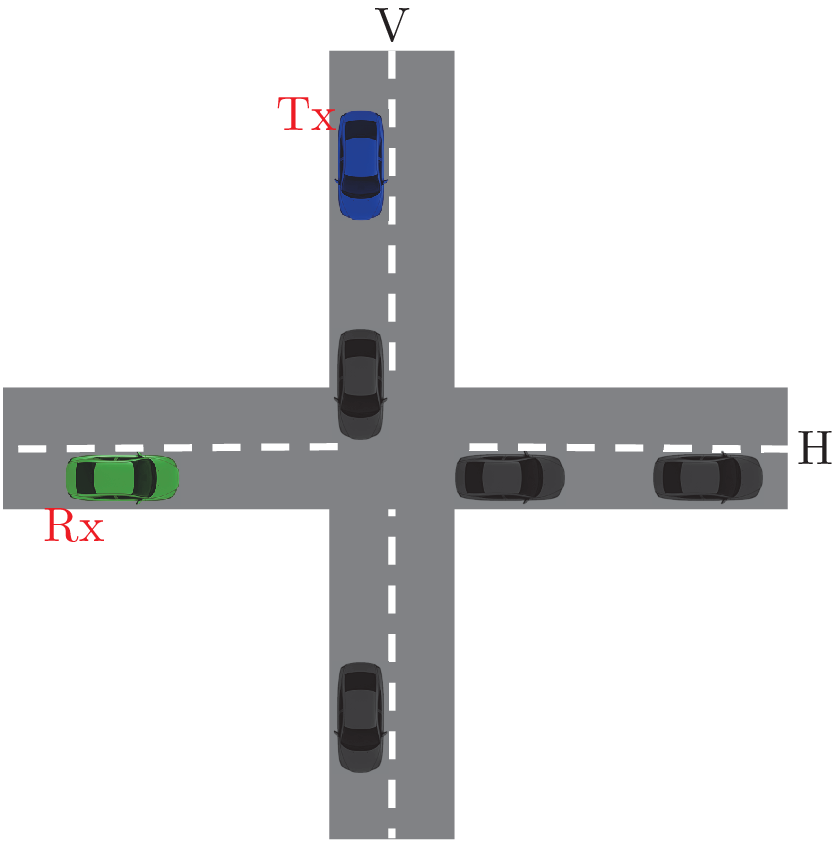}}\hfill\subfloat[abstraction]{\includegraphics[width=0.4\columnwidth]{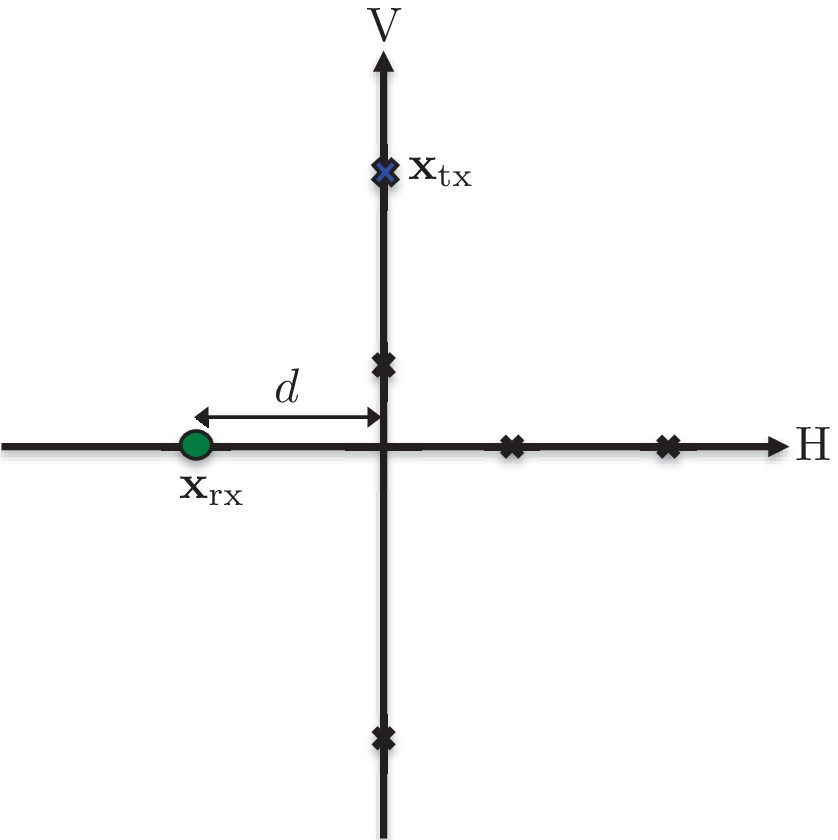}}

\caption{\label{fig:SystemModel}Illustration of considered scenario: (a) A
two-way intersection scenario in which each road carries a stream
a vehicles, (b) the abstraction used in modeling. The Tx (indicated
by the blue car) can be at any location, while the target Rx (green
car) is located on road ${\rm H}$. Other vehicles on the roads ${\rm H}$
and ${\rm V}$, of which some transmit concurrently and cause interference,
are shown as grey cars. }
 
\end{figure}

We consider an intersection scenario with two perpendicular roads,
as shown in Figure~\ref{fig:SystemModel}. For simplicity, we assume
that the width of the two roads indicated by ${\rm H}$ and ${\rm V}$
can be neglected, and that the roads each carry a stream of vehicles,
modeled as one-dimensional homogeneous Poisson point processes (PPPs).
The intensity of vehicles on both roads is denoted by $\lambda_{{\rm H}}$
and $\lambda_{{\rm V}}$, and the point processes describing the location
of the vehicles on the two roads are represented by $\Phi_{{\rm H}}\sim\text{PPP}(\lambda_{{\rm H}})$
and $\Phi_{{\rm V}}\sim\text{PPP}(\lambda_{{\rm V}})$. The positions
of individual vehicles (also referred to as nodes) on the two roads
${\rm H}$ and ${\rm V}$ are denoted by $\mathbf{x}_{i}=[x_{i},0]^{\mathrm{T}}$
and $\mathbf{x}_{i}=[0,y_{i}]^{\mathrm{T}}$, respectively, assuming
the roads are aligned with the horizontal and vertical axes. As both
vehicle-to-vehicle~(V2V) and infrastructure-to-vehicle~(I2V) communication
is of interest, we consider a transmitter~(Tx) with arbitrary location
$\mathbf{x}_{\mathrm{tx}}=[x_{\mathrm{tx}},y_{\mathrm{tx}}]^{\mathrm{T}}$.\footnote{Note that in the case the Tx is on one of the roads, it can belong
to either $\Phi_{{\rm H}}$ or $\Phi_{{\rm V}}$ (but does not necessarily
have to) as the results still hold due to Slivnyak's Theorem \cite[Theorem A.5]{Haenggi2008}} The Tx broadcasts with a fixed transmission power $P$. Without loss
of generality, we consider a receiver (Rx) on the ${\rm H}$-road
at location $\mathbf{x}_{\mathrm{rx}}=[x_{\mathrm{rx}},0]^{\mathrm{T}}$,
i.e., at a distance $d=\left|x_{\mathrm{rx}}\right|$ away from the
the intersection.\footnote{Note that due to the symmetry of the scenario this also captures the
case when Rx is on the V-road } The signal propagation comprises power fading $S$ and path loss
$l(\mathbf{x}_{\mathrm{tx}},\mathbf{x}_{\mathrm{rx}})$. At the Rx,
the signal is further affected by white Gaussian noise with noise
power $N$ and interference from other concurrently transmitting vehicles
on the H- and V-road. The amount of interference experienced by the
Rx depends on the choice of MAC protocol. For a given MAC scheme,
the position of interfering vehicles at a given time can be represented
by the thinned point processes $\Phi_{{\rm H}}^{{\rm MAC}}$ and $\Phi_{{\rm V}}^{{\rm MAC}}$.\footnote{For a general MAC scheme, the thinning process is inhomogeneous.}
We can express the signal-to-interference-plus-noise ratio~(SINR)
as
\begin{align}
 & {\rm SINR}=\frac{P\,S_{0}l(\mathbf{x}_{\mathrm{tx}},\mathbf{x}_{\mathrm{rx}})}{\sum_{\mathbf{x}\in\Phi_{{\rm H}}^{{\rm MAC}}}P\,S_{\mathbf{x}}l(\mathbf{x},\mathbf{x}_{\mathrm{rx}})+\sum_{\mathbf{x}\in\Phi_{{\rm V}}^{{\rm MAC}}}P\,S_{\mathbf{x}}l(\mathbf{x},\mathbf{x}_{\mathrm{rx}})+N},\label{eq:SINR}
\end{align}
where $S_{0}$ denotes the fading on the useful link and $S_{\mathbf{x}}$
denotes the fading on an interfering link for an interferer at location
$\mathbf{x}$. A packet is considered to be successfully received
if the SINR exceeds a threshold $\beta$. 

Our aim is to analytically characterize (i) the probability that the
Rx successfully receives a packet sent by the Tx; (ii) the throughput
of the link between Tx and Rx. This problem is challenging due to
the specific propagation conditions and interference levels experienced
in these intersection scenarios. In the next section, we will describe
these in more detail. 
\begin{rem}
While the scenario considered here is simple, it can easily be extended
to cases where the width of the roads can not be ignored (without
introducing significant modeling errors) by splitting the road into
several lanes, each modeled as a new road/lane, as discussed in Section~\ref{sub:Rural_Intersection}.
Furthermore, general multi road/lane extensions can be used to explicitly
model interference from other roads in the surrounding.
\end{rem}

\section{Models in Vehicular Communication\label{sec:Typical-Scenarios}}

Vehicular communication systems must be able to function in a large
variety of conditions, including in urban canyons and in rural settings.
In this section, we will discuss characteristics for vehicular channels
that are important from an SINR point of view, and detail different
models regarding path loss, fading, and MAC protocol.

\subsection{Power decay and blockage}

Extensive measurement campaigns \cite{Abbas2012,Karedal2011,Mecklenbrauker2011,Abbas2013,Mangel2011}
have been performed to characterize the vehicular channel in a variety
of propagation environments such as rural, highway, suburban, and
urban scenarios. As it is important to understand how the power decays
with distance, much efforts have been put into finding large-scale
path loss models, which characterize the slope of distance-dependent
power loss in decibels (dB). We will distinguish between line-of-sight
(LOS) and non-line-of-sight (NLOS) propagation, depending on whether
or not the direct LOS signal between a Rx and a Tx is blocked. For
LOS propagation, conventional path loss models, where power decays
approximately with the squared Euclidean distance between Rx and Tx
are well-accepted \cite{Mecklenbrauker2011}. For NLOS propagation,
e.g., in urban canyons, measurements indicate increased loss over
LOS propagation, with complex dependencies on the absolute position
of Tx and Rx, widths of the roads, and different loss exponents for
own and orthogonal road \cite{Abbas2013,Mangel2011}. The complexity
of these models render them intractable when it comes to mathematical
analysis, so we rely on the simpler and more tractable Manhattan model,
which was first proposed for modeling of similar scenarios in the
well-known WINNER II project \cite{WINNER2}. In particular, to allow
for mathematical analysis, we suggest the path loss of the following
form. 
\begin{itemize}
\item For NLOS propagation, where the direct line-of-sight~(LOS) between
the Rx and the Tx is blocked by buildings and the signals have to
propagate along the urban canyons formed by the orthogonal streets,
we use the Manhattan distance:
\begin{align}
 & l_{{\rm M}}(\mathbf{x}_{\mathrm{tx}},\mathbf{x}_{\mathrm{rx}})=A\left\Vert \mathbf{x}_{\mathrm{rx}}-\mathbf{x}_{\mathrm{tx}}\right\Vert _{1}^{-\alpha},\label{eq:ManhattanDist}
\end{align}
where $\left\Vert \mathbf{x}_{\mathrm{rx}}-\mathbf{x}_{\mathrm{tx}}\right\Vert _{1}$
is the $\ell_{1}$ norm, $\alpha>0$ is the path loss exponent, and
$A$ is a constant that depends on several factors such as antenna
characteristics, carrier frequency, and propagation environment. 
\item For LOS propagation, where the direct LOS between the Rx and the Tx
is unobstructed, we use the Euclidean distance 
\begin{align}
 & l_{{\rm E}}(\mathbf{x}_{\mathrm{tx}},\mathbf{x}_{\mathrm{rx}})=A\left\Vert \mathbf{x}_{\mathrm{rx}}-\mathbf{x}_{\mathrm{tx}}\right\Vert _{2}^{-\alpha},\label{eq:euclideanDist}
\end{align}
where $\left\Vert \mathbf{x}_{\mathrm{rx}}-\mathbf{x}_{\mathrm{tx}}\right\Vert _{2}$
is the $\ell_{2}$ norm. Note that the values of $\alpha$ and $A$
might be different in (\ref{eq:euclideanDist}) and (\ref{eq:ManhattanDist}).
\end{itemize}

\subsection{Random power variations due to fading \label{sub:Fading-distributions}}

Fading refers to random fluctuations in the received power around
the average received power, given by the path loss. The fading experienced
on a link depends on the scenario and the environment, and is typically
modeled as a random variable \cite{WildemeerschSR14}. For example,
near a rural intersection, vehicles are likely to communicate via
LOS links, and exponential fading is considered an appropriate model
\cite{Cheng2007,Mangel2011}. On the contrary, if the intersection
is located in an urban environment with tall buildings, the fading
for NLOS links is modeled using a log-normal model \cite{Mangel2011,Abbas2013},
with typical values on power variations with respect to the path loss
for NLOS intersections are in the range of 3--6 dB. Based on these
empirical results, we will model the fading as log-normal (and approximated
by an Erlang random variable for mathematical tractability -- see
Section~\ref{sub:MGF-of-fading}) for NLOS links and as exponential
for LOS links.

\subsection{MAC protocols}

The MAC protocol governs when a user can access the channel, and aims
to control the interference in the network. The two most common MAC
protocols for ad-hoc networks are Aloha and CSMA. In Aloha, which
is the simpler of the two, nodes that have a packet to send, access
the channel  during a time slot with a probability $p\in[0,1]$. In
contrast, in CSMA, before sending a packet, a node verifies that the
channel is free by listening to the channel. Only if the channel is
free, the node transmits the packet. If the channel is busy, the node
is forced to wait a random back-off time before it can try again \cite{Sjoberg2013}.
The 802.11p standard, which has been designed for the first generation
vehicular networks, will rely on a CSMA/CA (collision avoidance) MAC.
We will consider CSMA as well as Aloha, as Aloha is easier to analyze
and has been argued to exhibit similar performance as CSMA, especially
for dense networks \cite{Blaszczyszyn2009}.

\section{Stochastic Geometry Analysis\label{sec:Stochastic-Geometry-Analysis}}

From Section~\ref{sec:Typical-Scenarios}, it is apparent that vehicular
communication systems will operate under a variety of propagation
conditions. In this section, we describe a general and unified methodology
to compute the communication performance for all these conditions,
as well as different MAC protocols. In particular, we will determine
(i) the packet reception probability $\mathbb{P}(\beta,\mathbf{x}_{\mathrm{rx}},\mathbf{x}_{\mathrm{tx}})$,
i.e., the probability that a receiver located at $\mathbf{x}_{\mathrm{rx}}$
can successfully decode a transmission from a transmitter located
at $\mathbf{x}_{\mathrm{tx}}$, in the presence of interferers on
the H- and V-road; (ii) the throughput ${\rm \mathcal{T}}(\beta,\mathbf{x}_{\mathrm{rx}},\mathbf{x}_{\mathrm{tx}})$,
i.e., the expected rate for the link between the Rx and Tx at locations
$\mathbf{x}_{\mathrm{rx}}$ and $\mathbf{x}_{\mathrm{tx}}$, accounting
for both the packet reception probability and the probability of gaining
access to the channel. Both $\mathbb{P}(\beta,\mathbf{x}_{\mathrm{rx}},\mathbf{x}_{\mathrm{tx}})$
and ${\rm \mathcal{T}}(\beta,\mathbf{x}_{\mathrm{rx}},\mathbf{x}_{\mathrm{tx}})$
depend on the loss function, fading distribution, and the MAC protocol.
Note that the loss function and fading distribution relate to the
power decay and blockage as well as the random signal variations in
the specific scenario, while the MAC protocol relates to number of
interferers and their locations. Several applications of this methodology
will be discussed in Section~\ref{sec:Case-Studies}.

\subsection{Packet reception probability }

To derive the packet reception probability for the intersection scenario,
we start by accounting for the fading distribution of the useful link.
We express 
\begin{align}
 & \mathbb{P}(\beta,\mathbf{x}_{\mathrm{rx}},\mathbf{x}_{\mathrm{tx}})=\Pr({\rm SINR}\geq\beta)\nonumber \\
 & =\Pr\left(S_{0}\geq\left(I_{{\rm H}}+I_{{\rm V}}+\tilde{N}\right)\beta/l(\mathbf{x}_{\mathrm{tx}},\mathbf{x}_{\mathrm{rx}})\right)
\end{align}
in which $\tilde{N}=N/P$ and $I_{{\rm H}}=\sum_{\mathbf{x}\in\Phi_{{\rm {\rm H}}}^{{\rm MAC}}}S_{\mathbf{x}}l(\mathbf{x},\mathbf{x}_{\mathrm{rx}})$
while $I_{{\rm V}}=\sum_{\mathbf{x}\in\Phi_{{\rm V}}^{{\rm MAC}}}S_{\mathbf{x}}l(\mathbf{x},\mathbf{x}_{\mathrm{rx}})$.
Conditioning on the path loss, we can now write the packet reception
probability as 
\begin{align}
 & \mathbb{P}(\beta,\mathbf{x}_{\mathrm{rx}},\mathbf{x}_{\mathrm{tx}})\label{eq:PsuccStep1}\\
 & =\mathbb{E}_{I_{\mathrm{H}},I_{\mathrm{{\rm V}}}}\left\{ \bar{F}_{S_{0}}\left(\left(I_{\mathrm{H}}+I_{\mathrm{V}}+\tilde{N}\right)\beta/l(\mathbf{x}_{\mathrm{tx}},\mathbf{x}_{\mathrm{rx}})\right)\right\} \nonumber \\
 & =\iint\bar{F}_{S_{0}}\left(\left(t_{1}+t_{2}+\tilde{N}\right)\tilde{\beta}\right)f_{I_{{\rm H}},I_{{\rm V}}}(t_{1},t_{2}){\rm d}t_{1}{\rm d}t_{2},\nonumber 
\end{align}
where $\tilde{\beta}=\beta/l(\mathbf{x}_{\mathrm{tx}},\mathbf{x}_{\mathrm{rx}})$
and $\bar{F}_{S_{o}}(s_{0})$ is the complementary cumulative distribution
function~(CCDF) of the random variable $S_{0}$, evaluated in $s_{0}$. 

The expression (\ref{eq:PsuccStep1}) can be interpreted in two ways:
(i) as the expectation of $\bar{F}_{S_{0}}((I_{\mathrm{H}}+I_{\mathrm{V}}+\tilde{N})\beta/l(\mathbf{x}_{\mathrm{tx}},\mathbf{x}_{\mathrm{rx}}))$
with respect to the interference distribution; and (ii) as the transformation
of the interference distribution with a kernel function determined
by the CCDF of the fading distribution of the useful link. In either
interpretation, the distributions of the interference and the fading
play an important role. Note that for all relevant fading distributions
of the useful link, (\ref{eq:PsuccStep1}) will result in the Laplace
transform~(LT) of the interference distribution or a function of
LTs of the interference distribution. It is therefore convenient to
express these distributions through their (LT) or, equivalently, their
moment generating function~(MGF).

\subsubsection{LT of the interference}

From (\ref{eq:PsuccStep1}), we see that the packet reception probability
$\mathbb{P}(\beta,\mathbf{x}_{\mathrm{rx}},\mathbf{x}_{\mathrm{tx}})$
is a function of the interference distribution, which itself depends
on the location of the Rx and the interferers, as well as their fading
distributions and path loss. For a general MAC protocol the interference
from the H- and V-road are not independent. However, for the MAC protocols
studied in this paper the interference distribution factorizes as
$f_{I_{{\rm H}},I_{{\rm V}}}(t_{1},t_{2})=f_{I_{{\rm H}}}(t_{1})f_{I_{{\rm V}}}(t_{2})$.
In fact, the interference is independently thinned on the H- and V-road
in the case of Aloha, while for the CSMA scheme we can approximate
the joint interference distribution as the product of the marginals,
where the dependence is captured by a location dependent thinning
of the original PPPs \cite{Haenggi2011}. This means that the interfering
point processes $\Phi_{{\rm H}}^{{\rm MAC}}$ and $\Phi_{{\rm V}}^{{\rm MAC}}$
either are, or are approximated as PPPs (for more details see Section~\ref{sub:Intensity-of-thePPP}),
and that we can focus on a single road ${\rm R}\in\{{\rm H},V\}$,
with interference distribution $f_{I_{{\rm R}}}$. The Laplace transform
of $f_{I_{{\rm R}}}$ is defined as 
\begin{equation}
\mathcal{L}_{I_{{\rm R}}}(s)=\mathbb{E}[\exp(-sI_{{\rm R}})],\label{eq:MGF1}
\end{equation}
in which 
\begin{equation}
I_{{\rm R}}=\sum_{\mathbf{x}\in\Phi_{{\rm R}}^{{\rm MAC}}}S_{\mathbf{x}}l(\mathbf{x},\mathbf{x}_{\mathrm{rx}}).\label{eq:inferferencegeneric}
\end{equation}
Substitution of (\ref{eq:inferferencegeneric}) into (\ref{eq:MGF1})
then yields
\begin{align}
\mathcal{L}_{I_{{\rm R}}}(s) & \overset{\left(a\right)}{=}\mathbb{E}_{\Phi}\left[\prod_{\mathbf{x}\in\Phi_{{\rm R}}^{{\rm MAC}}}\mathbb{E}_{S_{\mathbf{x}}}\left\{ \exp\left(-s\,S_{\mathbf{x}}l(\mathbf{x},\mathbf{x}_{\mathrm{rx}})\right)\right\} \right]\\
 & =\mathbb{E}_{\Phi}\left[\prod_{\mathbf{x}\in\Phi_{{\rm R}}^{{\rm MAC}}}\mathcal{L}_{S_{\mathbf{x}}}\left(s\,l(\mathbf{x},\mathbf{x}_{\mathrm{rx}})\right)\right]\\
 & \overset{\left(b\right)}{=}\exp\!\left(\!\!-\!\!\int_{-\infty}^{+\infty}\!\lambda_{{\rm R}}^{{\rm MAC}}\left(\mathbf{x}\left(z\right),\mathbf{x}_{\mathrm{tx}}\right)\left(1-\mathcal{L}_{S_{\mathbf{x}}}\left(s\,l(\mathbf{x}(z),\mathbf{x}_{\mathrm{rx}})\right)\right){\rm d}z\!\right),\label{eq:LT_General_Integral-1}
\end{align}
where (a) holds due to the independence of the fading parameters,
$\mathbb{E}_{\Phi}\left[\cdot\right]$ is the expectation operator
with respect to the location of the interferers, and $\mathcal{L}_{S_{\mathbf{x}}}\left(\cdot\right)$
is the LT of the fading distribution of the interfering link; (b)
is due to the probability generating functional~(PGFL) for a PPP
\cite[Definition A.5]{Haenggi2008}, in which $\lambda_{{\rm R}}^{{\rm MAC}}\left(\mathbf{x}(z),\mathbf{x}_{\mathrm{tx}}\right)$
represents the intensity of the PPP $\Phi_{{\rm R}}^{{\rm MAC}}$,
which depends on the specific MAC protocol and in some cases on the
transmitter's location. Note that in (\ref{eq:LT_General_Integral-1}),
the intensity is defined over $z\in\mathbb{R}$, which represents
the position along the road ${\rm R}\in\{{\rm H},V\}$, where 
\begin{equation}
\mathbf{x}\left(z\right)=\left\{ \begin{array}{cc}
[z\,0]^{\mathrm{T}} & ,{\rm R}={\rm H}\\{}
[0\,z]^{\mathrm{T}} & ,{\rm R}={\rm V}
\end{array}\right..
\end{equation}
To determine $\mathcal{L}_{I_{{\rm R}}}(s)$, we must be able to compute
the integral (\ref{eq:LT_General_Integral-1}), which involves knowledge
of $\lambda_{\text{{\rm R}}}^{{\rm MAC}}\left(\mathbf{x}\left(z\right),\mathbf{x}_{\mathrm{tx}}\right)$
and $\mathcal{L}_{S_{\mathbf{x}}}\left(s\right)$. 
\begin{rem}
The Laplace transform of the interference can also be computed using
the principle of stochastic equivalence \cite{Blaszczyszyn2013},
where the LT in case of an arbitrary fading distribution can be found
based on the LT in case of Rayleigh fading, given an appropriate scaling
of the system parameters. 
\end{rem}

\subsubsection{LT of fading\label{sub:MGF-of-fading}}

For many relevant fading distributions, the LT is known, including
for exponential, Gamma, Erlang, and $\chi^{2}$ random variables.
While the log-normal distribution is harder to deal with, it can be
approximated by the Erlang distribution \cite{Abou-Rjeily2010}, which
combines tractability with expressiveness. When $S_{\mathbf{x}}\sim\mathrm{E}\left(k,\theta\right)$,
i.e., an Erlang distribution with shape parameter $k\in\mathbb{N}$
and rate parameter $1/\theta>0$, then 
\begin{equation}
\mathcal{L}_{S_{\mathbf{x}}}\left(s\right)=\left(1+s\theta\right)^{k}.
\end{equation}
As special cases, (i) $k=1$ corresponds to an exponential distribution
with mean $\theta$; (ii) $\theta=1/k$ corresponds to Nakagami-m
power fading.
\begin{rem}
When the fading of the useful link is exponentially distributed, (\ref{eq:PsuccStep1})
allows us to interpret $\mathbb{P}(\beta,\mathbf{x}_{\mathrm{rx}},\mathbf{x}_{\mathrm{tx}})$
as the LT of the interference, so that $\mathbb{P}(\beta,\mathbf{x}_{\mathrm{rx}},\mathbf{x}_{\mathrm{tx}})$$=\exp\left(-\tilde{N}\tilde{\beta}/\theta\right)$$\mathcal{L}_{I_{{\rm H}}}(\tilde{\beta}/\theta)$$\mathcal{L}_{I_{{\rm V}}}(\tilde{\beta}/\theta).$
\end{rem}

\subsubsection{Intensity of the interfering PPPs\label{sub:Intensity-of-thePPP}}

The intensity $\lambda_{{\rm R}}^{{\rm MAC}}\left(\mathbf{x}\left(z\right),\mathbf{x}_{\mathrm{tx}}\right)$
of the interference depends on the type of MAC that is utilized. We
distinguish between two cases: Aloha with transmit probability $p\in[0,1]$,
and CSMA with contention region with radius $\delta\ge0$. 
\begin{itemize}
\item \textbf{Aloha:} For an Aloha MAC, the vehicles on each road will transmit
with a probability $p$. This leads to an independent thinning of
the PPPs, so that $\lambda_{{\rm R}}^{{\rm MAC}}\left(\mathbf{x}\left(z\right),\mathbf{x}_{\mathrm{tx}}\right)=p\lambda_{{\rm R}}$,
irrespective of $z$ or $\mathbf{x}_{\mathrm{tx}}$.
\item \textbf{CSMA:} For a CSMA MAC, a vehicle will transmit if it has the
lowest random timer within its sensing range (contention region).
This means that (i) the intensity is in this case also a function
of $\mathbf{x}_{{\rm tx}}$ as other nodes in its contention region
are forced to be silent when it is active; (ii) the interference from
the H- and V-road is not independent. The timer process and the corresponding
dependent thinning result in a Matérn hard-core process type II, which
can be approximated by a PPP with independently thinned node density.
The approximation of the hard-core process by a PPP is shown to be
accurate in \cite{Haenggi2011} and has been applied in the context
of heterogeneous cellular networks, for instance in \cite{Cho2013}.\footnote{The extension to CSMA schemes with discrete back-off timers has been
proposed in \cite{Tong}, which retains concurrent transmitters due
to the non-zero probability of nodes with the same timer value.}  When the transmitter at ${\bf x}_{{\rm tx}}$ is active the resulting
intensity of the PPPs used to approximate the point process of interferers
can be expressed as
\begin{equation}
\lambda_{{\rm R}}^{{\rm MAC}}\left(\mathbf{x}\left(z\right),{\bf x}_{{\rm tx}}\right)=\begin{cases}
\begin{array}[t]{l}
p_{A}\left(\mathbf{x}\left(z\right)\right)\lambda_{{\rm R}}\\
0
\end{array} & \begin{array}[t]{l}
\left\Vert \mathbf{x}\left(z\right)-\mathbf{{\bf x}_{{\rm tx}}}\right\Vert >\delta\\
\left\Vert \mathbf{x}\left(z\right)-\mathbf{{\bf x}_{{\rm tx}}}\right\Vert \leq\delta
\end{array}\end{cases}.\label{eq:Apparent_Intensity_CSMA}
\end{equation}
In (\ref{eq:Apparent_Intensity_CSMA}), $p_{A}\left(\mathbf{x}\left(z\right)\right)$
is the access probability of a node. The access probability (which
is used to thin the original process) is the probability that the
given node has the smallest random timer in the corresponding contention
region (in this case modeled as a 2-dimensional ball $\mathcal{B}_{2}(\mathbf{x}\left(z\right),\delta)$
with radius $\delta$ centered at location $\mathbf{x}\left(z\right)$),
and can for one of the roads be expressed as
\begin{eqnarray}
p_{A}(\mathbf{x}\left(z\right)) & = & \int_{0}^{1}\exp(-t\Lambda(\mathcal{B}_{2}(\mathbf{x}\left(z\right),\delta)))\mathrm{d}t\\
 & = & \frac{1-\exp(-\Lambda(\mathcal{B}_{2}(\mathbf{x}\left(z\right),\delta)))}{\Lambda(\mathcal{B}_{2}(\mathbf{x}\left(z\right),\delta))},\label{eq:CSMA_Pa}
\end{eqnarray}
where 
\begin{equation}
\Lambda(\mathcal{B}_{2}(\mathbf{x}\left(z\right),\delta))=\begin{cases}
2\delta\lambda_{{\rm R}} & \left\Vert \mathbf{x}\left(z\right)\right\Vert >\delta\\
2\delta\lambda_{{\rm R}}+2\sqrt{\delta^{2}-\left\Vert \mathbf{x}\left(z\right)\right\Vert ^{2}}\lambda_{{\rm R'}} & \left\Vert \mathbf{x}\left(z\right)\right\Vert \leq\delta
\end{cases}
\end{equation}
represents the average number of nodes in the contention region. Note
that the average number of nodes, and thus the access probability
depends on the position $z$ along the road and the intensities $\lambda_{{\rm R}}$
and $\lambda_{{\rm R'}}$, which here represent the intensities of
the unthinned processes on the relevant road $\mathrm{R}$ and the
other road, respectively. 
\end{itemize}

\subsection{Throughput}

From a system perspective, the packet reception probability is not
sufficient to characterize the performance, since a MAC that allows
few concurrent transmissions leads to high packet reception probabilities
but low throughputs. Thus, to be able to compare the impact of different
MAC protocols, we characterize the throughput for the intersection
scenario, i.e., the number of bits transmitted per unit time and bandwidth
on a specific link. For the general case with a receiver and transmitter
located at $\mathbf{x}_{\mathrm{rx}}$ and $\mathbf{x}_{\mathrm{tx}}$,
respectively, we express the throughput as 

\begin{equation}
\mathcal{T}\left(\beta,\mathbf{x}_{\mathrm{rx}},\mathbf{x}_{\mathrm{tx}}\right)=p_{A}(\mathbf{x}_{\mathrm{tx}})\mathbb{P}(\beta,\mathbf{x}_{\mathrm{rx}},\mathbf{x}_{\mathrm{tx}})\log_{2}\left(1+\beta\right)\label{eq:Link-Throughput}
\end{equation}
where $p_{A}(\mathbf{x}_{\mathrm{tx}})$ is the access probability
of a transmitter located at $\mathbf{x}_{\mathrm{tx}}$, i.e., the
probability that the transmitter obtains access to the channel to
transmit a packet. For the Aloha MAC, the access probability is simply
$p_{A}(\mathbf{x}_{\mathrm{tx}})=p$, while for the CSMA case the
access probability is given in (\ref{eq:CSMA_Pa}) and depends on
the void probability in the 2-dimensional ball used to model the contention
region around $\mathbf{x}_{\mathrm{tx}}$.

\subsection{General Procedure \label{sub:General-expression}}

Given the analysis in the previous subsections, the general procedure
for determining the packet reception probability $\mathbb{P}(\beta,\mathbf{x}_{\mathrm{rx}},\mathbf{x}_{\mathrm{tx}})$
and the throughput ${\rm \mathcal{T}}(\beta,\mathbf{x}_{\mathrm{rx}},\mathbf{x}_{\mathrm{tx}})$
is thus as follows: 
\begin{itemize}
\item \textbf{Step 1: }Determine the fading LT $\mathcal{L}_{S_{\mathbf{x}}}\left(s\right)$
for the interfering links, as described in Section~\ref{sub:MGF-of-fading}. 
\item \textbf{Step 2: }Determine the intensity of the interference PPP $\lambda_{{\rm R}}^{{\rm MAC}}\left(\mathbf{x}\left(z\right),{\bf x}_{{\rm tx}}\right)$
for ${\rm R\in\{H,V\}}$, as described in Section~\ref{sub:Intensity-of-thePPP}. 
\item \textbf{Step 3: }From step 1 and step 2, determine the LT of the interference
$\mathcal{L}_{I_{{\rm R}}}(s)$ for ${\rm R\in\{H,V\}}$ using (\ref{eq:LT_General_Integral-1}). 
\item \textbf{Step 4:} Determine the fading LT $\mathcal{L}_{S_{0}}\left(s\right)$
for the useful link, as described in Section~\ref{sub:MGF-of-fading}. 
\item \textbf{Step 5:} From step 4 and step 3, determine $\mathbb{P}(\beta,\mathbf{x}_{\mathrm{rx}},\mathbf{x}_{\mathrm{tx}})$
using (\ref{eq:PsuccStep1}), either by drawing samples from the interference,
or by considering the CCDF of the fading on the useful link as a kernel
in a transformation (i.e., evaluating a function of LTs of the interference
distribution). Finally, use the obtained packet reception probability
$\mathbb{P}(\beta,\mathbf{x}_{\mathrm{rx}},\mathbf{x}_{\mathrm{tx}})$
in conjunction with the access probability $p_{A}(\mathbf{x}_{\mathrm{tx}})$
used in step 2 to determine the throughput ${\rm \mathcal{T}}(\beta,\mathbf{x}_{\mathrm{rx}},\mathbf{x}_{\mathrm{tx}})$.
\end{itemize}
Whether or not each step is tractable depends on the assumptions we
make regarding the loss function, the fading distribution, and the
MAC protocol, which will be further discussed in Section~\ref{sec:Case-Studies}.

\section{Case Studies\label{sec:Case-Studies}}

In this Section we present three case studies to show how the different
models presented in the paper can be used to model both rural and
urban intersection scenarios, and how shadowing, LOS blockage, and
different MAC protocols affect the performance of the communication
system. In Case~I, we present the most basic case which corresponds
to the rural setting, while in Case~II, we show how an urban intersection
can be modeled. Finally, in Case~III, we will study the impact of
the different MAC protocols. In each case study, we will discuss the
tractability of the resulting expressions, validate modeling assumptions
through simulations, and provide numerical performance examples.

\subsection{Case I - Rural intersection with Aloha\label{sub:Rural_Intersection}}

In the rural intersection scenario \cite{steinmetz2014communication,Steinmetz2015},
vehicles are assumed to communicate via LOS links. Hence, path loss
is described by the Euclidean distance loss function $l_{{\rm E}}(\cdot)$,
defined in (\ref{eq:euclideanDist}) with path loss exponent $\alpha=2$,
while power fading is modeled with an exponential distribution (i.e.,
$S\sim E\left[1,1\right]$), for both useful and interfering links.
Furthermore, we consider an Aloha MAC with transmit probability~$p$.

\subsubsection{Packet reception probability }

Using the procedure from Section~\ref{sub:General-expression}, the
packet reception probability for the rural intersection scenario is
given in Proposition~\ref{prop:Rural_Euclidean_And_Exponential}
(see also \cite{steinmetz2014communication,Steinmetz2015}). 
\begin{prop}
\label{prop:Rural_Euclidean_And_Exponential}Given a slotted Aloha
MAC with transmit probability\textcolor{blue}{~}$p$, exponential
fading\textcolor{blue}{~}(i.e, $S\sim E(1,1)$) for each link, Euclidean
loss function\textcolor{blue}{~}$l_{{\rm E}}(\cdot)$ with path loss
exponent $\alpha=2$, and a scenario as outlined in Section~\ref{sec:System-Model},
the packet reception probability can be expressed as

\textup{
\begin{align}
 & \mathbb{P}(\beta,\mathbf{x}_{\mathrm{rx}},\mathbf{x}_{\mathrm{tx}})=\exp\left(-\frac{N\beta\left\Vert \mathbf{x}_{{\rm rx}}-\mathbf{x_{{\rm tx}}}\right\Vert _{2}^{2}}{PA}\right)\label{eq:caseIexpression}\\
 & \times\exp\left(-p\lambda_{{\rm H}}\pi\sqrt{\beta}\left\Vert \mathbf{x}_{{\rm rx}}-\mathbf{x_{{\rm tx}}}\right\Vert _{2}\right)\exp\left(-\frac{p\lambda_{{\rm V}}\pi\beta\left\Vert \mathbf{x}_{{\rm rx}}-\mathbf{x_{{\rm tx}}}\right\Vert _{2}^{2}}{\sqrt{\beta\left\Vert \mathbf{x}_{{\rm rx}}-\mathbf{x_{{\rm tx}}}\right\Vert _{2}^{2}+d^{2}}}\right)\nonumber 
\end{align}
}\end{prop}
\begin{IEEEproof}
See Appendix~\ref{sec:Proof-of-Proposition-1}. 
\end{IEEEproof}
We note that the packet reception probability comprises three factors:
the first factor corresponds to the packet reception probability in
the absence of interferers; the second factor captures the reduction
of the packet reception probability due to interferers on the H-road;
the third factor captures the additional reduction of packet reception
probability due to interferers on the V-road. 
\begin{rem}
Proposition~\ref{prop:Rural_Euclidean_And_Exponential} can be extended
in a number of ways:\end{rem}
\begin{itemize}
\item As was noted in \cite{Steinmetz2015} additional roads/lanes with
arbitrary orientations can be accounted for, each road contributing
with an additional factor to the packet reception probability. This
approach can for example be used to take into account interference
from surrounding roads. Furthermore, it can be used to handle cases
where the width of the roads can no longer be ignored, by splitting
the road into several lanes.
\item Extensions to scenarios with non-homogeneous PPPs are also possible,
in order to model, e.g., clustering of vehicles due to traffic congestions.
In general this requires numeric integration to evaluate the LTs of
the interference distribution, but for special cases such as piecewise
linear intensity functions, closed-form expressions can be found.
\end{itemize}

\subsubsection{Numerical results}

\begin{figure}
\begin{centering}
\includegraphics[width=0.6\columnwidth]{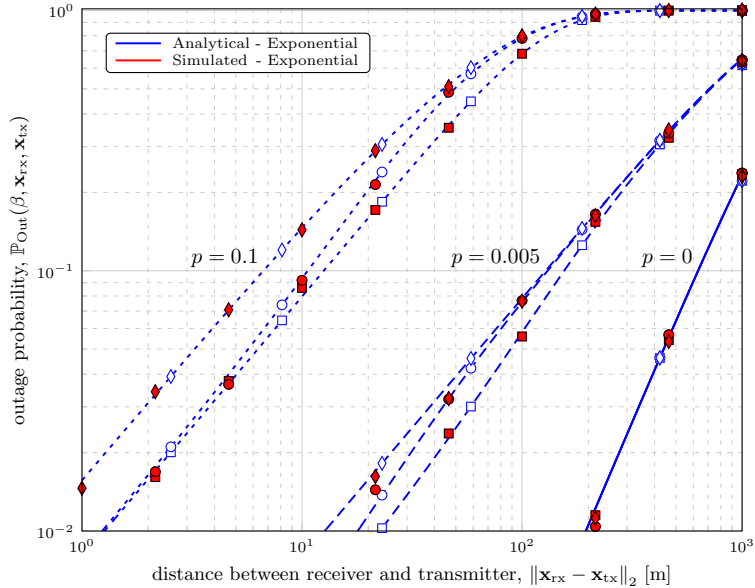}
\par\end{centering}

\caption{\label{fig:Outage_Prob_Case_I}Comparison of analytical and simulated
outage probability $\mathbb{P}_{{\rm Out}}(\beta,\mathbf{x}_{\mathrm{rx}},\mathbf{x}_{\mathrm{tx}})$
versus distance between transmitter and receiver $\left\Vert \mathbf{x}_{{\rm rx}}-\mathbf{x_{{\rm tx}}}\right\Vert _{2}$
for different distances to the intersection $d$ as well as different
transmit probabilities $p$. The distances $d$ are 0 m (diamonds),
100 m (circles) and 500 m (squares). }
\end{figure}

Throughout the remainder of the paper, we consider an intersection
where the intensity of vehicles on the two roads are $\lambda_{{\rm H}}=\lambda_{{\rm V}}=0.01$
(i.e., with an average inter-vehicle distance of 100 m). Furthermore,
we assume a noise power $N$ of $-99\,\mathrm{dBm}$, an SINR threshold
of $\beta=8\,\mathrm{dB}$ \cite{Sjoberg2013}, and that $A=3\cdot10^{-5}$,
approximately matching the conditions in \cite{Karedal2011}. We set
the transmit power to $P=100\,\mathrm{mW}$, corresponding to 20 dBm.
For the purpose of visualization, we show the \emph{outage probability}
$\mathbb{P}_{{\rm Out}}(\beta,\mathbf{x}_{\mathrm{rx}},\mathbf{x}_{\mathrm{tx}})=1-\mathbb{P}(\beta,\mathbf{x}_{\mathrm{rx}},\mathbf{x}_{\mathrm{tx}})$
instead of the packet reception probability. Figure~\ref{fig:Outage_Prob_Case_I}
shows the analytical outage probability for the rural intersection
scenario as a function of distance between transmitter and receiver
$\left\Vert \mathbf{x}_{{\rm rx}}-\mathbf{x_{{\rm tx}}}\right\Vert _{2}$
for different distances to the intersection $d\in\{0\text{ m},100\text{ m},500\text{ m}\}$
and different transmit probabilities $p\in\{0,0.005,0.1\}$. We observe
that the outage probability increases with the distance between the
receiver and the transmitter, and that interference has a negative
impact on the performance as the outage probability is higher for
increased transmit probabilities. In the absence of interferers ($p=0)$
the system achieves an outage probability of 10\textcolor{magenta}{~}\%
when the receiver and transmitter are spaced approximately 600 m apart.
When $p$ is increased to $0.005$ the communication range is drastically
reduced to about 130\textcolor{magenta}{~}m, due to the interference.
Furthermore, the figure reflects the location dependence of the outage
probability with respect to the intersection, and we can see that
the outage probability increases when the receiver is closer to the
intersection. For the purpose of validation, we have added Monte Carlo
simulation with 10,000 realizations of the PPPs and fading parameters,
perfectly matching the analytical expressions. 

\subsection{Case II - Urban Intersection with Aloha}

This case study, model an urban intersection scenario, with the Tx
on the V-road and the Rx on the H-road. Signals arriving to the Rx
from the V-road are assumed to be in NLOS, modeled through Manhattan
path loss and Erlang fading (which serves as an approximation of log-normal
fading). Signals arriving to the Rx from the own H-road are in LOS,
modeled through Euclidean path loss and exponential fading.

\subsubsection{Packet reception probability}

The packet reception probability for the urban intersection scenario
is given in Proposition~\ref{prop:Urban_Manhattan_And_Erlang}. 
\begin{prop}
\label{prop:Urban_Manhattan_And_Erlang}Given a slotted Aloha MAC
with transmit probability~$p$, Erlang fading~(i.e., $S\sim E(k_{0},\theta_{0})$)
and Manhattan loss function~$l_{{\rm M}}(\cdot)$ for the useful
link, Erlang fading (i.e., $S\sim E(k_{V},\theta_{V})$) and Manhattan
loss function for the interfering links from the V-road, exponential
fading~(i.e, $S\sim E(1,1)$) and Euclidean loss function~$l_{{\rm E}}(\cdot)$
for the interfering links from the H-road, and a scenario as outlined
in Section~\ref{sec:System-Model}, the packet reception probability
can be expressed as
\begin{equation}
\mathbb{P}(\beta,\mathbf{x}_{\mathrm{rx}},\mathbf{x}_{\mathrm{tx}})\!=\!e^{-\frac{\zeta N}{P}}\sum_{i=0}^{k_{0}-1}\sum_{j=0}^{i}\!\!\binom{i}{j}\frac{\zeta^{i}}{i!}\!C^{(j)}D^{(i,j)},
\end{equation}
where
\begin{equation}
C^{(j)}=\sum_{n=0}^{j}\binom{j}{n}\left(\frac{N}{P}\right){}^{j-n}\left(-1\right)^{n}e^{-\kappa\sqrt{\zeta}}\zeta^{-n}\sum_{l=0}^{n}\sum_{m=0}^{l}\frac{\left(-1\right)^{m}\left(-\kappa\sqrt{\zeta}\right)^{l}\left(\frac{2-m+l-2n}{2}\right)_{n}}{m!\left(-m+l\right)!},
\end{equation}
and
\begin{eqnarray}
 &  & D^{(i,j)}=\\
 &  & \left(-1\right)^{i-j}\frac{d^{i-j}}{d(\zeta)^{i-j}}\exp\left(-2p\lambda_{{\rm V}}\sum_{q=0}^{k_{{\rm V}}-1}\binom{k_{{\rm V}}}{q}\frac{1}{\alpha\Gamma\left[k_{{\rm V}}\right]}\left(\frac{A\zeta}{\theta_{{\rm V}}}\right)^{-q}\Gamma\left[\frac{1}{\alpha}+q\right]\right.\nonumber \\
 &  & \times\!\left.\!\left(\!\!-\!\left(\frac{A\zeta}{\theta_{{\rm V}}}\right)^{-\frac{1}{\alpha}+q}\!\Gamma\!\left[-\frac{1}{\alpha}+k_{{\rm V}}-q\right]\!+\!d^{1+\alpha q}\Gamma\!\left[k_{{\rm V}}\right]{}_{2}F_{1}\!\left[k_{{\rm V}},\frac{1}{\alpha}\!+\!q,1\!+\!\frac{1}{\alpha}+q,-\frac{d^{\alpha}}{A\zeta\theta_{{\rm V}}}\right]\!\right)\!\right)\!.\nonumber 
\end{eqnarray}
In which $\kappa=2p\lambda_{{\rm H}}A^{1/\text{\ensuremath{\alpha}}}\pi/\alpha\csc\left(\pi/\alpha\right)$
and $\zeta=\beta\left\Vert \mathbf{x}_{{\rm rx}}-\mathbf{x}_{{\rm tx}}\right\Vert _{1}^{\alpha}/\left(A\theta_{0}\right).$\end{prop}
\begin{IEEEproof}
See Appendix~\ref{sec:Proof-of-Proposition-2}. 
\end{IEEEproof}
We observe that the analytical expressions become more involved when
changing the loss function as well as the fading distribution for
the links to the V-road, but in contrast to the rural intersection
scenario it is possible to obtain closed form expressions for a general
$\alpha$ (this is because Manhattan path loss for the interferers
from the V-road is easier to handle than Euclidean path loss). Furthermore,
it should be noted that if the Tx is assumed to be on the H-road,
the expressions become more compact. Moreover, similarly as for the
model presented in \cite{Mangel2011}, Proposition~\ref{prop:Urban_Manhattan_And_Erlang}
only gives realistic results when the Rx and the Tx are at least a
few meters away from the intersection. This is because when the Rx
is at the intersection, all links become LOS, while when the Tx is
at the intersection, the useful link becomes LOS. In either case,
the corresponding links should be modeled with exponential fading,
rather than Erlang fading.

\subsubsection{Numerical results}

\begin{figure}
\begin{centering}
\includegraphics[width=0.6\columnwidth]{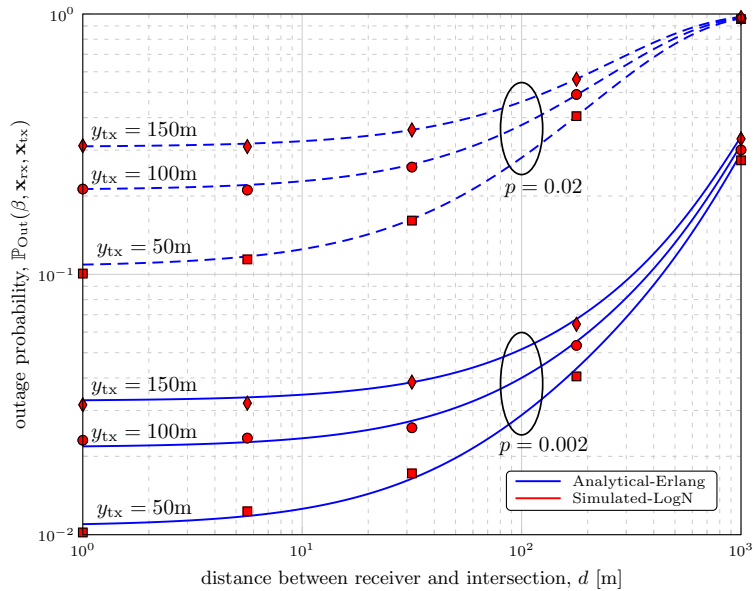}
\par\end{centering}

\caption{\label{fig:Outage_Prob_Urban_Case_II-1}Comparison between analytical
and simulated outage probability $\mathbb{P}_{{\rm Out}}(\beta,\mathbf{x}_{\mathrm{rx}},\mathbf{x}_{\mathrm{tx}})$
versus distance between receiver and intersection $d$ for different
Tx positions and transmit probabilities. Lines show analytical results
based on Proposition\textcolor{blue}{~\ref{prop:Urban_Manhattan_And_Erlang}},
while markers show simulations for the same scenario but with 3.2
dB log-normal fading. The different transmit probabilities\textcolor{blue}{~}$p$,
which the outage probability is plotted for are 0.002\textcolor{blue}{~}(solid)
and 0.02\textcolor{blue}{~}(dashed) lines. }
\end{figure}

In this section we intend to validate the accuracy of the Erlang approximation.
We consider the same parameters for the LOS propagation as in Section~\ref{sub:Rural_Intersection}.
We set the Aloha transmit probability to $p\in\{0.002,0.02\}$. For
all NLOS links, we use the same value of $A$ as in the LOS links,
set $\alpha=2$ , and consider the fading to be log-normal with 3.2
dB standard deviation, as in one of the intersections studied in \cite{Abbas2013}.
Maximum likelihood fitting of the Erlang distribution to the log-normal
distribution yielded $k_{0}=k_{{\rm V}}=k=2$ and $\theta_{0}=\theta_{{\rm V}}=\theta\approx0.66$.
In Figure~\ref{fig:Outage_Prob_Urban_Case_II-1}, a comparison between
the outage probability obtained by evaluating Proposition~\ref{prop:Urban_Manhattan_And_Erlang}
under the Erlang approximation, and Monte Carlo simulations for the
same scenario but with 3.2 dB log-normal shadowing is shown. First,
we observe that the analytical results based on the Erlang approximation
agrees well with the simulations, i.e., in terms of outage probability
the Erlang fading provides a good approximation to the actual log-normal
fading. Furthermore, we see that as expected, lower transmit probability,
as well as placing the Tx closer to the intersection results in lower
outage probabilities. Even though the results shows very good agreement
between the analytical results and the simulations, it should be mentioned
that the approximation become less accurate when the standard deviation
increases. In particular, this is noticeable when the standard deviation
of the log-normal distribution exceeds 3.8 dB, as the Erlang distribution
obtained from the fitting then reverts to the exponential distribution.
Furthermore, when decreasing the standard deviation of the log-normal
distribution, the estimated value of $k$ rapidly increases, so that
the model becomes less tractable. 

\subsection{Case III - Aloha vs CSMA}

In this final case study, we will focus on the MAC protocol and how
it affects performance and tractability. To do this, we start from
the rural intersection scenario, but replace the Aloha MAC with a
CSMA MAC. As the MAC affects not only the packet reception probability
but also the access probability, we will also consider throughput
in this case study.

\subsubsection{Packet reception probability\label{sub:Analytical-results-MAC}}

The packet reception probability for the CSMA case is given in Proposition~\ref{prop:CSMA_MAC}. 
\begin{prop}
\label{prop:CSMA_MAC}Given a CSMA MAC with contention radius~$\delta$,
exponential fading~(i.e, $S\sim E(1,1)$) for each link, Euclidean
loss function~$l_{{\rm E}}(\cdot)$ with path loss exponent $\alpha=2$,
and a scenario as outlined in Section~\ref{sec:System-Model}, the
success probability can be expressed as
\begin{align}
 & \mathbb{P}(\beta,\mathbf{x}_{\mathrm{rx}},\mathbf{x}_{\mathrm{tx}})=e^{-\frac{N\tilde{\beta}}{P}}\mathcal{L}_{I_{{\rm H}}}(\tilde{\beta})\mathcal{L}_{I_{{\rm V}}}(\tilde{\beta}),\label{eq:SucessProb_CSMA}
\end{align}
where $\tilde{\beta}=\beta/l_{{\rm E}}(\mathbf{x}_{\mathrm{tx}},\mathbf{x}_{\mathrm{rx}})$,
and 
\begin{align}
\mathcal{L}_{I_{{\rm H}}}(s) & =\exp\left(-\int_{-\infty}^{+\infty}\frac{\lambda_{{\rm H}}^{{\rm MAC}}\left(\left[x,0\right]^{\mathrm{T}},{\bf x}_{{\rm tx}}\right)}{1+\left|x_{{\rm rx}}-x\right|^{2}/As}{\rm d}x\right)
\end{align}

\begin{align}
\mathcal{L}_{I_{{\rm V}}}(s) & =\exp\left(-\int_{-\infty}^{+\infty}\frac{\lambda_{{\rm V}}^{{\rm MAC}}\left(\left[0,y\right]^{\mathrm{T}},{\bf x}_{{\rm tx}}\right)}{1+\bigl\Vert\left[x_{{\rm rx}},-y\right]^{\mathrm{T}}\bigr\Vert_{2}^{2}/As}{\rm d}y\right)
\end{align}
where $\lambda_{{\rm H}}^{{\rm MAC}}\left(\left[x,0\right]^{\mathrm{T}},{\bf x}_{{\rm tx}}\right)$
and $\lambda_{{\rm V}}^{{\rm MAC}}\left(\left[0,y\right]^{\mathrm{T}},{\bf x}_{{\rm tx}}\right)$
are given in (\ref{eq:CSMA_Intensity_X}) and (\ref{eq:CSMA_Intensity_Y}),
respectively. \end{prop}
\begin{IEEEproof}
See Appendix~\ref{sec:Proof-of-Proposition-3}. 
\end{IEEEproof}
As can be seen from Proposition~\ref{prop:CSMA_MAC}, the expressions
we obtain still involve an integral that can be solved numerically
easily and efficiently. 

\begin{figure}
\begin{centering}
\includegraphics[width=0.6\columnwidth]{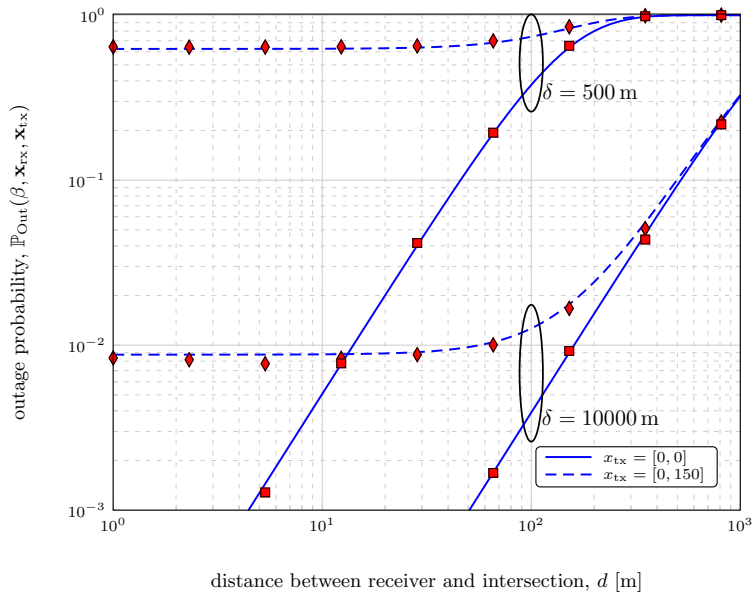}
\par\end{centering}

\caption{\label{fig:Case_III-3}Comparison of analytical (blue lines) and simulated
(red markers) outage probability $\mathbb{P}_{{\rm Out}}(\beta,\mathbf{x}_{\mathrm{rx}},\mathbf{x}_{\mathrm{tx}})$
versus distance between receiver and intersection $d$, for different
transmitter locations ${\rm {\bf x}}_{{\rm tx}}$ as well as CSMA
contention radiuses $\delta$. The receiver is located on the H-road,
while the transmitter location is fixed to either ${\bf x}_{{\rm tx}}=[0,0\text{]}$
(solid lines) or ${\bf x}_{{\rm tx}}=[0,150\text{]}$ (dashed lines).
The different CSMA contention radiuses are $\delta=500\,\mathrm{m}$
and $\delta=10000\,\mathrm{m}$, which in the region where the access
probability is constant, i.e., far away from the intersection, corresponds
to $p_{A}=0.1$ and $p_{A}=0.005$, respectively. }
\end{figure}

\subsubsection{Numerical results}

 In order to evaluate the accuracy of the approximation introduced
in Section~\ref{sub:Intensity-of-thePPP}, we start by comparing
the analytically calculated outage probability to a simulation with
50000 realizations of the fading parameters and the hard-core process
induced by the dependent thinning resulting from the CSMA scheme.
This comparison can be seen in Figure~\ref{fig:Case_III-3}, which
shows the analytical and simulated outage probability as a function
of the distance between the receiver and the intersection for two
different transmitter locations (${\bf x}_{{\rm tx}}=[0,0\text{]}$
and ${\bf x}_{{\rm tx}}=[0,150\text{]}$), as well as two different
CSMA contention radiuses $\delta\in\{500\,\mathrm{m},10000\,\mathrm{m}\}$.
 We observe good correspondence between simulation and analytical
results, and an increase in outage probability with increased distance
to the intersection. We also note that when ${\bf x}_{{\rm tx}}=[0,0\text{]}$,
it is possible to compare Figure \ref{fig:Case_III-3} with Figure
\ref{fig:Outage_Prob_Case_I}. We note that for $\delta=10000\,\mathrm{m}$,
for a distance of 100 m between Rx and intersection, CSMA has an outage
probability of 0.003, while Aloha is over 25 times worse, with an
outage probability of 0.08. 

\begin{figure}
\begin{centering}
\includegraphics[width=0.6\columnwidth]{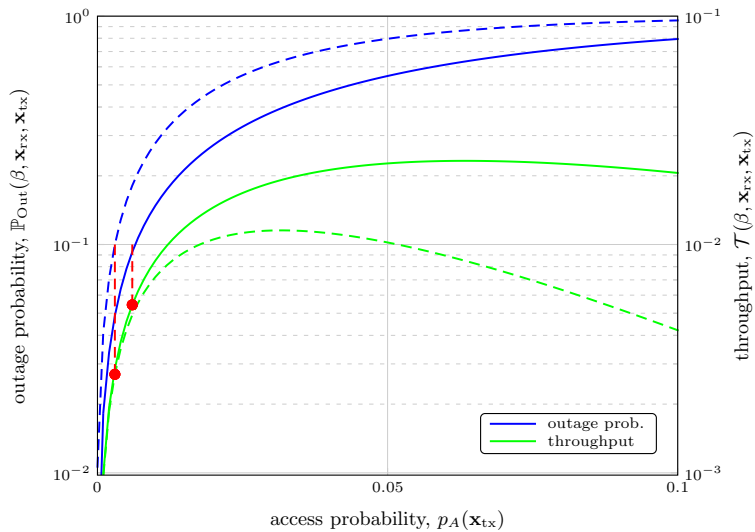}
\par\end{centering}

\caption{\label{fig:Case_III-1}Outage probability $\mathbb{P}_{{\rm Out}}(\beta,\mathbf{x}_{\mathrm{rx}},\mathbf{x}_{\mathrm{tx}})$
and throughput $\mathcal{T}\left(\beta,\mathbf{x}_{\mathrm{rx}},\mathbf{x}_{\mathrm{tx}}\right)$
for the Aloha case as a function of the transmitter access probability
$p_{A}({\rm \mathbf{x}_{{\rm tx}})}$. The receiver is located at
$\mathbf{x}_{{\rm rx}}=[0,0\text{]}$, and solid lines correspond
to $R_{\mathrm{comm}}=100\,\mathrm{m}$, while dashed lines correspond
to $R_{\mathrm{comm}}=200\,\mathrm{m}$. The red circles indicate
the maximum throughput that is possible to achieve while guaranteeing
that the outage probability is kept below the target value of 10~\%. }
\end{figure}

\begin{figure}
\begin{centering}
\includegraphics[width=0.6\columnwidth]{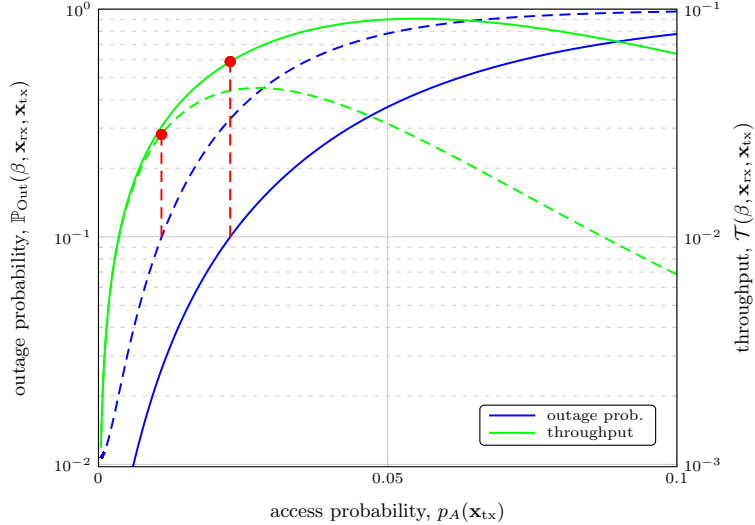}
\par\end{centering}

\caption{\label{fig:Case_III-2}Outage probability $\mathbb{P}_{{\rm Out}}(\beta,\mathbf{x}_{\mathrm{rx}},\mathbf{x}_{\mathrm{tx}})$
and throughput $\mathcal{T}\left(\beta,\mathbf{x}_{\mathrm{rx}},\mathbf{x}_{\mathrm{tx}}\right)$
for the CSMA case as a function of the transmitter access probability
$p_{A}({\rm {\bf x}_{{\rm tx}}})$. The receiver is located at $\mathbf{x}_{{\rm rx}}=[-100,0\text{]}$,
and solid lines correspond to $R_{\mathrm{comm}}=100\,\mathrm{m}$,
while dashed lines correspond to $R_{\mathrm{comm}}=200\,\mathrm{m}$.
The red circles indicate the maximum throughput that is possible to
achieve while guaranteeing that the outage probability is kept below
the target outage probability of 10~\%. }
\end{figure}

To further study the performance gains achieved by using CSMA compared
to Aloha, we now look at both outage probability and throughput for
a specific receiver and transmitter configuration. The configuration
that we consider is $\mathbf{x}_{\mathrm{rx}}=[0\,0]^{\mathrm{T}}$
and $\mathbf{x}_{\mathrm{tx}}=[R_{\mathrm{comm}}\,0]^{\mathrm{T}}$.
Note that for the Aloha case this placement results in the worst possible
throughput for a fixed $l_{{\rm E}}(\mathbf{x}_{\mathrm{tx}},\mathbf{x}_{\mathrm{rx}})$.
Figure~\ref{fig:Case_III-1} and Figure~\ref{fig:Case_III-2}, show
the outage probability as well as throughput as a function of the
access probability~$p_{A}({\rm {\bf x}_{{\rm tx}}})$, for two different
values on $R_{{\rm comm}}\in\{100\,\mathrm{m},200\,\mathrm{m}\}$.

For Aloha (Figure~\ref{fig:Case_III-1}), we see that with an increase
in $p_{A}({\rm {\bf x}_{{\rm tx}}})$, outage probability increases
due to the presence of more interferers. The throughput first increases
(due to more active transmitters) and then decreases (due to overwhelming
amounts of interference), leading to an optimal value of $p_{A}({\rm {\bf x}_{{\rm tx}}})$.
However, in order to guarantee a certain quality of service, one must
also consider a guarantee on the outage probability. For instance,
if we want to guarantee an outage probability of less than 10~\%
on the link when $R_{\mathrm{comm}}=100\,\mathrm{m}$, the optimal
value of $p_{A}({\rm {\bf x}_{{\rm tx}}})\approx0.006$, leading to
a throughput of around 0.0055 bits per unit time and bandwidth.

For CSMA (Figure~\ref{fig:Case_III-2}), a low access probability
(i.e., large contention region) reduces the outage probability. Similar
to Aloha, the throughput first increases with increased access probability
and then decreases. To achieve an outage probability below 10~\%
when $R_{\mathrm{comm}}=100\,\mathrm{m}$, the optimal value of $p_{A}({\rm {\bf x}_{{\rm tx}}})\thickapprox0.023$
(corresponding to a contention radius~$\delta$ of about 1100~m),
results in a throughput of about 0.059 bits per unit time and bandwidth.
Hence, in this scenario, using CSMA instead of Aloha leads to more
than a tenfold increase in the throughput for the same communication
range.

\section{Conclusions\label{sec:Conclusions}}

We have provided an overview of the dominant propagation properties
of vehicular communication systems near intersections, for both rural
and urban scenarios. Based on these properties, we proposed a general
procedure to analytically determine packet reception probabilities
of individual transmissions as well throughput, mainly applicable
to 802.11p communication. In contrast to traditional cellular networks,
the one-dimensional road geometry leads to non-homogeneous packet
reception probabilities and throughputs. 

We have applied and validated this procedure to three case studies,
relevant for vehicular applications. The results indicate that the
procedure is sufficiently general and flexible to deal with a variety
of scenarios, that its performance results match well with simulations,
and that it is able to capture the clustering of vehicles close to
the intersection. The procedure can thus serve as a useful design
tool for communication system engineers, complementing simulations
and experiments. 

Possible avenues for future research includes validation of the model
agains actual measurements, adoption of advanced MAC schemes as well
as 5G D2D features.

\appendices{}

\section{Proof of Proposition 4\label{sec:Proof-of-Proposition-1}}

In order to determine the packet reception probability when $S\sim E(1,1)$,
we follow the general procedure from Section~\ref{sub:General-expression}. 

\textbf{Step 1: }The fading LT for the interfering links can be expressed
as $\mathcal{L}_{S_{\mathbf{x}}}\left(s\right)=1/(1+s)$.

\textbf{Step 2:} According to Section~\ref{sub:Intensity-of-thePPP}
the intensity of the two interfering PPPs $\Phi_{{\rm H}}^{{\rm MAC}}$
and $\Phi_{{\rm V}}^{{\rm MAC}}$ are $\lambda_{{\rm H}}^{{\rm MAC}}\left([x\,0]^{\mathrm{T}},\mathbf{x}_{\mathrm{tx}}\right)=p\lambda_{{\rm H}}$
and $\lambda_{{\rm V}}^{{\rm MAC}}\left([0,y]^{\mathrm{T}},\mathbf{x}_{\mathrm{tx}}\right)=p\lambda_{{\rm V}}$,
respectively. 

\textbf{Step 3:} The LT of the interference for the two roads are
derived as follows\textbf{\textit{. }}Since the fading of the interfering
links is exponentially distributed, (\ref{eq:LT_General_Integral-1})
simplifies to 
\begin{equation}
\mathcal{L}_{I_{{\rm R}}}(s)=\exp\left(-\int_{-\infty}^{+\infty}\frac{\lambda_{{\rm R}}^{{\rm MAC}}\left(\mathbf{x}\left(z\right),\mathbf{x}_{\mathrm{tx}}\right)}{1+1/(s\theta l(\mathbf{x}\left(z\right),\mathbf{x}_{\mathrm{rx}}))}{\rm d}z\right).\label{eq: MGF_of_Interference_Exp}
\end{equation}
Using (\ref{eq: MGF_of_Interference_Exp}) for the horizontal road
with Euclidean path loss, and bearing in mind that $\mathbf{x}(z)=[z\,0]^{\mathrm{T}}$
, we can write 
\begin{align}
 & \mathcal{L}_{I_{{\rm {\rm H}}}}(s)=\exp\left(-\int_{-\infty}^{+\infty}\frac{\lambda_{{\rm H}}^{{\rm MAC}}\left(\mathbf{x}(z),\mathbf{x}_{\mathrm{tx}}\right)}{1+\left\Vert \mathbf{x}_{{\rm rx}}-\mathbf{x}(z)\right\Vert _{2}^{\alpha}/As}{\rm d}z\right)\\
 & \overset{(a)}{=}\exp\left(-p\lambda_{{\rm H}}\int_{-\infty}^{+\infty}\frac{1}{1+\left|x_{{\rm rx}}-x\right|^{\alpha}/As}{\rm d}x\right)\\
 & \overset{(b)}{=}\exp\left(-2p\lambda_{{\rm H}}\left(As\right)^{1/\text{\ensuremath{\alpha}}}\int_{0}^{+\infty}\frac{1}{1+u^{\alpha}}{\rm d}u\right)\\
 & =\exp\left(-2p\lambda_{{\rm H}}\left(As\right)^{1/\text{\ensuremath{\alpha}}}\pi/\alpha\csc\left(\pi/\alpha\right)\right)\label{eq:Ref_General_Alpha}
\end{align}
where $(a)$ uses the fact that the intensity is $p\lambda_{{\rm H}}$
on the H-road, and $(b)$ involves a change of variable $u=\left|x_{{\rm rx}}-x\right|/(As)^{1/\alpha}$.
For the particular case of $\alpha=2$ the LT of the interference
further simplifies to
\begin{align}
\mathcal{L}_{I_{{\rm {\rm H}}}}(s) & =\exp\left(-p\lambda_{{\rm H}}\pi\sqrt{As}\right).\label{eq:LTX_Exp}
\end{align}
For the V-road, using (\ref{eq: MGF_of_Interference_Exp}), we can
in a similar way as for the H-road write
\begin{align}
\mathcal{L}_{I_{{\rm {\rm V}}}}(s) & =\exp\left(-p\lambda_{{\rm V}}\int_{-\infty}^{+\infty}\frac{1}{1+\left\Vert \left[x_{{\rm rx}},-y\right]^{T}\right\Vert _{2}^{\alpha}/As}{\rm d}y\right).
\end{align}
Now using that the distance $\left\Vert \left[x_{{\rm rx}},-y\right]^{T}\right\Vert _{2}=\sqrt{x_{{\rm rx}}^{2}+y^{2}}=\sqrt{d^{2}+y^{2}}$
we can introduce $r_{y}=\sqrt{d^{2}+y^{2}}$, with $dr_{y}/dy=y/r_{y}$.
Noting that a PPP remains PPP under a non-linear transformation according
to the mapping theorem \cite[Theorem A.1]{Haenggi2008}, we have
\begin{align}
 & \mathcal{L}_{I_{{\rm {\rm V}}}}(s)=\exp\!\left(\!\!-2p\lambda_{{\rm V}}\int_{d}^{+\infty}\!\!\frac{r_{y}}{\sqrt{r_{y}^{2}-d^{2}}\left(1+r_{y}^{\alpha}/As\right)}{\rm d}r_{y}\!\right)\\
 & =\exp\!\left(\!\!-p\lambda_{{\rm V}}\left(As\right)^{1/\alpha}\!\!\int_{\omega_{0}}^{+\infty}\!\!\frac{1}{\sqrt{\omega-\omega_{0}}\left(1+\omega^{\alpha/2}\right)}{\rm d}\omega\!\right)
\end{align}
where we have carried out the following change of variable $\omega=\left(r_{y}/\left(As\right)^{1/\text{\ensuremath{\alpha}}}\right)^{2}$,
and further introduced $\omega_{0}=\left(d/\left(As\right)^{1/\alpha}\right)^{2}$.
For $\alpha=2$, the integral can be computed as $\int_{\omega_{0}}^{+\infty}(\sqrt{\omega-\omega_{0}}\left(1+\omega\right))^{-1}{\rm d}\omega$$=\text{\ensuremath{\pi}/\ensuremath{\sqrt{1+\omega_{0}}}}$,
which yields
\begin{align}
 & \mathcal{L}_{I_{{\rm V}}}\left(s\right)=\exp\left(\frac{-p\lambda_{{\rm V}}\pi As}{\sqrt{As+d^{2}}}\right)\label{eq:LTY_Exp}
\end{align}
Note that for $d\to0$, (\ref{eq:LTY_Exp}) reverts to (\ref{eq:LTX_Exp}),
while for $d\to+\infty$, (\ref{eq:LTY_Exp}) tends to one. 

\textbf{Step 4:} The fading on the useful link is characterized by
its LT $\mathcal{L}_{S_{0}}\left(s\right)=1/(1+s)$ and CCDF $\bar{F}_{S_{0}}(s)=\exp\left(-s\right).$

\textbf{Step 5:} Using the LT of the interference from Step 3, and
the CCDF of the fading from Step 4 as a kernel, we can now determine
$\mathbb{P}(\beta,\mathbf{x}_{\mathrm{rx}},\mathbf{x}_{\mathrm{tx}})$
through (\ref{eq:PsuccStep1}). First using the CCDF, and evaluating
it in the desired point, we can write
\begin{align}
 & \bar{F}_{S_{0}}\left(\left(t_{1}+t_{2}+\tilde{N}\right)\tilde{\beta}\right)=\exp\left(-\left(t_{1}+t_{2}+\tilde{N}\right)\tilde{\beta}\right)\label{eq:CCDF_Desired_Exp}
\end{align}
As the interference from the H- and V-road is independent (i.e., $\Phi_{{\rm H}}^{{\rm MAC}}$
and $\Phi_{{\rm V}}^{{\rm MAC}}$ are independent) we can now use
(\ref{eq:CCDF_Desired_Exp}) to express the transform in (\ref{eq:PsuccStep1})
as
\begin{align}
 & \mathbb{P}(\beta,\mathbf{x}_{\mathrm{rx}},\mathbf{x}_{\mathrm{tx}})\nonumber \\
 & =\exp\left(-\tilde{N}\tilde{\beta}\right)\int_{0}^{+\infty}f_{I_{{\rm H}}}(t_{1})\exp\left(-t_{1}\tilde{\beta}\right){\rm d}t_{1}\int_{0}^{+\infty}f_{I_{{\rm V}}}(t_{2})\exp\left(-t_{2}\tilde{\beta}\right){\rm d}t_{2}\\
 & =e^{-\tilde{N}\tilde{\beta}}\mathcal{L}_{I_{{\rm H}}}\left(\tilde{\beta}\right)\mathcal{L}_{I_{{\rm V}}}\left(\tilde{\beta}\right)
\end{align}
Using the results from step 3, and the variable changes $\tilde{\beta}=\beta/l_{{\rm E}}(\mathbf{x}_{\mathrm{tx}},\mathbf{x}_{\mathrm{rx}})$
and $\tilde{N}=N/P$, finally allow us to express the packet reception
probability as (\ref{eq:caseIexpression}).

\section{Proof of Proposition 6 \label{sec:Proof-of-Proposition-2}}

We use the procedure from Section~\ref{sub:General-expression}.

\textbf{Step 1: }The fading LTs for the interfering links from the
H-road and the V-road can be expressed as $\mathcal{L}_{S_{\mathbf{x}}}\left(s\right)=1/(1+s)$
and $\mathcal{L}_{S_{{\rm x}}}(s)=1/(1+s\theta_{V})^{k_{V}}$, respectively.

\textbf{Step 2: }According to Section~\ref{sub:Intensity-of-thePPP}
the intensity of the two PPPs $\Phi_{{\rm H}}^{{\rm MAC}}$ and $\Phi_{{\rm V}}^{{\rm MAC}}$
are $p\lambda_{{\rm H}}$ and $p\lambda_{{\rm V}}$, respectively. 

\textbf{Step 3: }The LT of the interference for the two roads are
derived in the following way. For the H-road, with interferers ${\rm \mathbf{x}\in}\Phi_{{\rm H}}^{{\rm MAC}}$
, the fading LT as well as the loss function are the same as in the
rural intersection case. Thus we can according to (\ref{eq:Ref_General_Alpha})
expresse the LT of the interference for a general $\alpha$ as 
\begin{equation}
\mathcal{L}_{I_{{\rm {\rm H}}}}(s)=\exp\left(-2p\lambda_{{\rm H}}\left(As\right)^{1/\text{\ensuremath{\alpha}}}\pi/\alpha\csc\left(\pi/\alpha\right)\right).
\end{equation}
For the V-road we now have fading LT $\mathcal{L}_{S_{{\rm x}}}(s)=1/(1+s\theta_{V})^{k_{V}}$,
intensity $p\lambda_{{\rm V}}$, and Manhattan loss function. Hence,
using (\ref{eq:LT_General_Integral-1}) we can write

\begin{eqnarray}
\mathcal{L}_{I_{V}}(s) & = & \exp\left(-\intop_{-\infty}^{\infty}\lambda_{{\rm V}}^{{\rm MAC}}\left(\mathbf{x}(z),\mathbf{x}_{\mathrm{tx}}\right)\left(1-\mathcal{L}_{S_{\mathbf{x}}}\left(s\,l_{{\rm {\rm M}}}(\mathbf{x}(z),\mathbf{x}_{\mathrm{rx}})\right)\right){\rm d}z\right)\\
 & = & \exp\left(-p\lambda_{{\rm V}}\intop_{-\infty}^{\infty}\left(1-\frac{1}{(1+s\theta_{V}A\left\Vert \left[x_{{\rm rx}},-y\right]^{T}\right\Vert _{1}^{-\alpha}){}^{k_{V}}}\right){\rm d}y\right)\\
 & \overset{(a)}{=} & \exp\left(-p\lambda_{{\rm V}}\sum_{q=0}^{k_{{\rm V}}-1}\binom{k_{{\rm V}}}{q}\intop_{-\infty}^{\infty}\frac{\left\Vert \left[x_{{\rm rx}},-y\right]^{T}\right\Vert _{1}^{\alpha q}b^{k_{{\rm V}}-q}}{(\left\Vert \left[x_{{\rm rx}},-y\right]^{T}\right\Vert _{1}^{\text{\ensuremath{\alpha}}}+b)^{k_{{\rm V}}}}{\rm d}y\right)\\
 & \overset{(b)}{=} & \exp\left(-p\lambda_{{\rm V}}\sum_{q=0}^{k_{{\rm V}}-1}\binom{k_{{\rm V}}}{q}\intop_{-\infty}^{\infty}\frac{(d+\left|y\right|)^{\alpha q}b^{k_{{\rm V}}-q}}{((d+\left|y\right|)^{\text{\ensuremath{\alpha}}}+b)^{k_{{\rm V}}}}{\rm d}y\right)\\
 & \overset{(c)}{=} & \exp\left(-p\lambda_{{\rm V}}\sum_{q=0}^{k_{{\rm V}}-1}\binom{k_{{\rm V}}}{q}\intop_{-\infty}^{\infty}\frac{u{}^{\alpha q}b^{k_{{\rm V}}-q}}{(u{}^{\text{\ensuremath{\alpha}}}+b)^{k_{{\rm V}}}}{\rm d}u\right)
\end{eqnarray}
where $(a)$ uses the Binomial Theorem and the variable change $s\theta_{{\rm V}}A\rightarrow b$,
$(b)$ uses that for points $\mathbf{x}\in\Phi_{{\rm V}}^{{\rm MAC}}$
the distance $\left\Vert \mathbf{x}_{{\rm rx}}-\mathbf{x}\right\Vert _{1}=\left|x_{{\rm rx}}\right|+\left|y\right|=d+\left|y\right|$,
and $(c)$ uses the variable change $d+\left|y\right|\rightarrow u$.
For $q\geq0$ , $k_{{\rm V}}\geq q+1$, $b\geq0$ and $d>0$ the integral
can be evaluated in closed form, and for a general $\alpha$ we can
express the LT of the interference as

\begin{eqnarray}
 &  & \mathcal{L}_{I_{V}}(s)=\\
 &  & \exp\left(-2p\lambda_{{\rm V}}\sum_{q=0}^{k_{{\rm V}}-1}\binom{k_{{\rm V}}}{q}\frac{1}{\alpha\Gamma\left[k_{{\rm V}}\right]}\left(\frac{As}{\theta_{{\rm V}}}\right)^{-q}\Gamma\left[\frac{1}{\alpha}+q\right]\right.\nonumber \\
 &  & \times\left.\left(-\left(\frac{As}{\theta_{{\rm V}}}\right)^{-\frac{1}{\alpha}+q}\Gamma\left[-\frac{1}{\alpha}+k_{{\rm V}}-q\right]+d^{1+\alpha q}\Gamma\left[k_{{\rm V}}\right]{}_{2}F_{1}\left[k_{{\rm V}},\frac{1}{\alpha}+q,1+\frac{1}{\alpha}+q,-\frac{d^{\alpha}}{As\theta_{{\rm V}}}\right]\right)\right),\nonumber 
\end{eqnarray}
where $_{2}F_{1}$ is the regularized hypergeometric function. Note
that for $\alpha=2$ and $k_{{\rm V}}=\theta_{{\rm V}}=1$ (i.e.,
exponential fading) this simplifies to

\begin{equation}
\mathcal{L}_{I_{{\rm {\rm V}}}}(s)=\exp\left(-p\lambda_{{\rm V}}\sqrt{As}\,\left(\pi-\text{2arctan\ensuremath{\left(\frac{d}{\sqrt{As}}\right)}}\right)\right),
\end{equation}
and when $d\rightarrow0$ we get 

\begin{equation}
\mathcal{L}_{I_{{\rm {\rm V}}}}(s)=\exp\left(-p\lambda_{{\rm V}}\pi\sqrt{As}\right),
\end{equation}
i.e., it reverts to the same form as $\mathcal{L}_{I_{{\rm {\rm H}}}}(s)$
in (\ref{eq:LTX_Exp}).

\textbf{Step 4: }The fading on the useful link is characterized by
its LT $\mathcal{L}_{S_{0}}\left(s\right)=1/(1+s\theta_{0})^{k_{0}}$
and CCDF
\begin{align}
\bar{F}_{S_{0}}(s) & =e^{-s/\theta_{0}}\sum_{i=0}^{k_{0}-1}\frac{1}{i!\theta_{0}^{i}}s^{i}\label{eq:CCDF_Erlang}
\end{align}

\textbf{Step 5: }In the same manner as in Appendix~\ref{sec:Proof-of-Proposition-1},
we now use the LTs of the interference from Step 3, and the CCDF of
the fading from Step 4 to determine $\mathbb{P}(\beta,\mathbf{x}_{\mathrm{rx}},\mathbf{x}_{\mathrm{tx}})$
through (\ref{eq:PsuccStep1}). First using the CCDF, and evaluating
it in the desired point, we can write 
\begin{align}
 & \bar{F}_{S_{0}}\left(\left(t_{1}+t_{2}+\tilde{N}\right)\tilde{\beta}\right)\\
 & =e^{-\text{\ensuremath{\tilde{\beta}}}\left(t_{1}+t_{2}+\tilde{N}\right)/\theta_{0}}\sum_{i=0}^{k_{0}-1}\frac{1}{i!\theta_{0}^{i}}\left(\tilde{\beta}\right)^{i}\left(t_{1}\!+\!t_{2}\!+\!\tilde{N}\right)^{i}\\
 & \overset{(a)}{=}e^{-\zeta\left(t_{1}+t_{2}+\tilde{N}\right)}\sum_{i=0}^{k_{0}-1}\frac{\zeta^{i}}{i!}\left(t_{1}\!+\!t_{2}\!+\!\tilde{N}\right)^{i}\\
 & \overset{(b)}{=}e^{-\zeta\tilde{N}}\!\sum_{i=0}^{k_{0}-1}\!\sum_{j=0}^{i}\!\!\binom{i}{j}\frac{\zeta^{i}}{i!}\!e^{-\zeta t_{1}}\!(\tilde{N}\!+\!t_{1})^{j}e^{-\zeta t_{2}}t_{2}^{i\!-\!j},\label{eq:CCDF_Desired_point-1-1}
\end{align}
where $(a)$ involves the variable change $\zeta=\tilde{\beta}/\theta_{0}$
and $(b)$ uses the Binomial Theorem. Due to the independence of the
interference we can now use (\ref{eq:CCDF_Desired_point-1-1}) to
express the transform in (\ref{eq:PsuccStep1}) as 
\begin{align}
\mathbb{P}(\beta,\mathbf{x}_{\mathrm{rx}},\mathbf{x}_{\mathrm{tx}}) & \!=\!e^{-\frac{\zeta N}{P}}\sum_{i=0}^{k_{0}-1}\sum_{j=0}^{i}\!\!\binom{i}{j}\frac{\zeta^{i}}{i!}\!C^{(j)}D^{(i,j)},\label{eq:SuccesProb_General-1-1}
\end{align}
where
\begin{align}
C^{(j)} & =\int_{0}^{+\infty}e^{-\zeta t_{1}}(\tilde{N}+t_{1})^{j}f_{I_{{\rm H}}}(t_{1}){\rm d}t_{1}\\
 & =\sum_{n=0}^{j}\binom{j}{n}\tilde{N}{}^{j-n}\mathcal{L}[t_{1}^{n}f_{I_{{\rm H}}}(t_{1})](\zeta)\\
 & =\sum_{n=0}^{j}\binom{j}{n}\left(\frac{N}{P}\right){}^{j-n}\left(-1\right)^{n}\frac{d^{n}}{d(\zeta)^{n}}\mathcal{L}_{I_{{\rm H}}}(\zeta)\label{eq:C3-1}
\end{align}
and
\begin{align}
D^{(i,j)} & =\int_{0}^{+\infty}e^{-\zeta t_{2}}t_{2}^{i-j}f_{I_{{\rm V}}}(t_{2}){\rm d}t_{2}\\
 & =\mathcal{L}[t_{2}^{i-j}f_{I_{{\rm V}}}(t_{2})](\zeta)\\
 & =\left(-1\right)^{i-j}\frac{d^{i-j}}{d(\zeta)^{i-j}}\mathcal{L}_{I_{{\rm V}}}(\zeta)
\end{align}
are obtained using the Laplace transform property $t^{n}f\left(t\right)\longleftrightarrow\left(-1\right)^{n}\frac{d^{n}}{d\zeta^{n}}\mathcal{L}\left[f\left(t\right)\right]\left(\zeta\right)$.
Note that (\ref{eq:SuccesProb_General-1-1}) and (\ref{eq:C3-1})
use the variable change $\tilde{N}=N/P$. Now using the results from
step 4 express the $n^{{\rm th}}$ derivative of the LT of the interference
from the H-road as 
\begin{equation}
\frac{d^{n}}{d(\zeta)^{n}}\mathcal{L}_{I_{{\rm H}}}(\zeta)=e^{-\kappa\sqrt{\zeta}}\zeta^{-n}\sum_{l=0}^{n}\sum_{m=0}^{l}\frac{\left(-1\right)^{m}\left(-\kappa\sqrt{\zeta}\right)^{l}\left(\frac{2-m+l-2n}{2}\right)_{n}}{m!\left(-m+l\right)!}
\end{equation}
where $\left(\cdot\right)_{n}$ is the Pochhammer symbol and $\kappa=2p\lambda_{{\rm H}}\left(A\right)^{1/\text{\ensuremath{\alpha}}}\pi/\alpha\csc\left(\pi/\alpha\right)$.
For the V-road, there is no general compact expression for the $n^{{\rm th}}$
derivative of $\mathcal{L}_{I_{{\rm V}}}(\zeta)$, but an explicit
expression can in principle be calculated for any $n$, $k_{{\rm V}}$
and $\theta_{{\rm V}}$.

\section{Proof of Proposition 7 \label{sec:Proof-of-Proposition-3}}

We use the procedure from Section~\ref{sub:General-expression}.

\textbf{Step 1: }The fading LT for the interfering links can be expressed
as $\mathcal{L}_{S_{\mathbf{x}}}\left(s\right)=1/(1+s)$.

\textbf{Step 2: }According to Section~\ref{sub:Intensity-of-thePPP},
the intensity of the two PPPs $\Phi_{{\rm H}}^{{\rm MAC}}$ and $\Phi_{{\rm V}}^{{\rm MAC}}$
are for this case also a function of the transmitter location $\mathbf{x}_{{\rm tx}}$.
Using (\ref{eq:Apparent_Intensity_CSMA}) we can express the intensity
for the H-road as
\begin{equation}
\lambda_{{\rm H}}^{{\rm MAC}}\left(\left[x,0\right]^{\mathrm{T}},{\bf x}_{{\rm tx}}\right)=\begin{cases}
\frac{1-\exp(-2\delta\lambda_{{\rm H}})}{2\delta} & x\in\text{\ensuremath{\mathcal{R}}}_{1}\\
\frac{1-\exp\left(-2\delta\lambda_{{\rm H}}-2\sqrt{\delta^{2}-x^{2}}\lambda_{{\rm V}}\right)\lambda_{{\rm H}}}{2\delta\lambda_{{\rm H}}+2\sqrt{\delta^{2}-x^{2}}\lambda_{{\rm V}}} & x\in\text{\ensuremath{\mathcal{R}}}_{2}\\
0 & \mathrm{else}
\end{cases}\label{eq:CSMA_Intensity_X}
\end{equation}
in which $\text{\ensuremath{\mathcal{R}}}_{1}=\{x|\left|x\right|>\delta\text{ and }\sqrt{(x-x_{\mathrm{tx}})^{2}+y_{\mathrm{tx}}^{2}}>\delta\}$
and $\text{\ensuremath{\mathcal{R}}}_{2}=\{x|\left|x\right|\le\delta\text{ and }\sqrt{(x-x_{\mathrm{tx}})^{2}+y_{\mathrm{tx}}^{2}}>\delta\}$.
Similarly for the V-road,
\begin{equation}
\lambda_{{\rm V}}^{{\rm MAC}}\left(\left[0,y\right]^{\mathrm{T}},{\bf x}_{{\rm tx}}\right)=\begin{cases}
\frac{1-\exp(-2\delta\lambda_{{\rm V}})}{2\delta} & y\in\text{\ensuremath{\mathcal{R}}}_{3}\\
\frac{1-\exp\left(-2\delta\lambda_{{\rm V}}-2\sqrt{\delta^{2}-y^{2}}\lambda_{{\rm {\rm H}}}\right)\lambda_{{\rm V}}}{2\delta\lambda_{{\rm V}}+2\sqrt{\delta^{2}-y^{2}}\lambda_{{\rm H}}} & y\in\text{\ensuremath{\mathcal{R}}}_{4}\\
0 & \mathrm{else}
\end{cases}\label{eq:CSMA_Intensity_Y}
\end{equation}
in which $\text{\ensuremath{\mathcal{R}}}_{3}=\{y|\left|y\right|>\delta\text{ and }\sqrt{(y-y_{\mathrm{tx}})^{2}+x_{\mathrm{tx}}^{2}}>\delta\}$
and $\text{\ensuremath{\mathcal{R}}}_{4}=\{y|\left|y\right|\le\delta\text{ and }\sqrt{(y-y_{\mathrm{tx}})^{2}+x_{\mathrm{tx}}^{2}}>\delta\}$. 

\textbf{Step 3: }Using (\ref{eq: MGF_of_Interference_Exp}) the LT
of the interference for the H- and V- road can be expresed as 
\begin{align}
 & \mathcal{L}_{I_{{\rm {\rm H}}}}(s)=\exp\left(\!-\!\int_{-\infty}^{+\infty}\frac{\lambda_{{\rm H}}^{{\rm MAC}}\left(\left[x,0\right]^{\mathrm{T}},{\bf x}_{{\rm tx}}\right)}{1+\left|x_{{\rm rx}}-x\right|^{\alpha}/As}{\rm d}x\!\right)\label{eq:CSMA_Int1}
\end{align}
and 
\begin{align}
 & \mathcal{L}_{I_{{\rm {\rm V}}}}(s)=\exp\left(\!-\!\int_{-\infty}^{+\infty}\frac{\lambda_{{\rm V}}^{{\rm MAC}}\left(\left[0,y\right]^{\mathrm{T}},{\bf x}_{{\rm tx}}\right)}{1+\left\Vert \left[x_{{\rm rx}},-y\right]^{\mathrm{T}}\right\Vert _{2}^{\alpha}/As}{\rm d}y\!\right)\label{eq:CSMA_Int2}
\end{align}

\textbf{Step 4: }The fading fading on the useful link is characterized
by its LT $\mathcal{L}_{S_{0}}\left(s\right)=1/(1+s)$ and CCDF $\bar{F}_{S_{0}}(s)=\exp\left(-s\right)$.

\textbf{Step 5: }By applying a location dependent thinning, we approximate
the interference from the H- and V-road as independent. Hence, as
the fading on the useful link is exponential (i.e., $S_{0}\sim E(1,1)$),
we can in the same way as in Appendix~\ref{sec:Proof-of-Proposition-1},
express the packet reception probability as $\mathbb{P}(\beta,\mathbf{x}_{\mathrm{rx}},\mathbf{x}_{\mathrm{tx}})=e^{-\tilde{N}\tilde{\beta}}\mathcal{L}_{I_{{\rm H}}}\left(\tilde{\beta}\right)\mathcal{L}_{I_{{\rm V}}}\left(\tilde{\beta}\right)$.
Using the results from Step 3, as well as the variable change $\tilde{N}=N/P$,
we can for the particular value of $\alpha=2$ finally obtain (\ref{eq:SucessProb_CSMA}).
Note that for a general transmitter location ${\bf x}_{{\rm tx}}$,
we are not able to evaluate the integrals in (\ref{eq:CSMA_Int1})
and (\ref{eq:CSMA_Int2}) in closed form, but have to resort to numerical
evaluation.

\bibliographystyle{IEEEtran}
\bibliography{Journal}

\end{document}